\documentclass[journal,twoside,web]{IEEEtran}
\usepackage[noadjust]{cite}
\usepackage{amsmath,amssymb,amsfonts}
\usepackage{algorithmic}
\usepackage{graphicx}
\usepackage{epstopdf}
\usepackage{textcomp}
\usepackage{amsmath}
\usepackage{graphicx}
\usepackage{cite}
\usepackage{multirow}
\usepackage{booktabs} 
\usepackage{subfigure}
\usepackage{marvosym}
\usepackage{float}
\usepackage{array}
\usepackage{color}
\usepackage{url}
\usepackage{pifont} % http://ctan.org/pkg/pifont
\begin{document}
\title{\huge{Bi-Modality Medical Image Synthesis Using Semi-Supervised Sequential Generative Adversarial Networks}}
\author{Xin Yang, Yi Lin*, Zhiwei Wang, Xin Li, Kwang-Ting Cheng \IEEEmembership{Fellow, IEEE} 
\thanks{The asterisk indicates the corresponding author.}
\thanks{Accepted by JBHI.}
}

\maketitle
% -----------------------------------------------------
\begin{abstract}
% ------------------------------------------
In this paper, we propose a bi-modality medical image synthesis approach based on sequential generative adversarial network (GAN) and semi-supervised learning. Our approach consists of two generative modules that synthesize images of the two modalities in a sequential order. A method for measuring the synthesis complexity is proposed to automatically determine the synthesis order in our sequential GAN. Images of the modality with a lower complexity are synthesized first, and the counterparts with a higher complexity are generated later. Our sequential GAN is trained end-to-end in a semi-supervised manner. In supervised training, the joint distribution of bi-modality images are learned from real paired images of the two modalities by explicitly minimizing the reconstruction losses between the real and synthetic images. To avoid overfitting limited training images, in unsupervised training, the marginal distribution of each modality is learned based on unpaired images by minimizing the Wasserstein distance between the distributions of real and fake images. We comprehensively evaluate the proposed model using two synthesis tasks based on three types of evaluate metrics and user studies. Visual and quantitative results demonstrate the superiority of our method to the state-of-the-art methods, and reasonable visual quality and clinical significance. Code is made publicly available at \url{https://github.com/hustlinyi/Multimodal-Medical-Image-Synthesis}.
% ------------------------------------------
\end{abstract}

% % -----------------------------------------------------
\section{Introduction}
\label{sec:introduction}
% ------------------------------------------
Multimodal medical imaging is widely considered to involve the incorporation of two or more imaging modalities, within the setting of a single examination using, for example, both functional and anatomical forms of magnetic resonance imaging (i.e., multiparametric MRI), or by performing ultrasound within MR, or x-ray computed tomography (CT) environment. Multimodal imaging could ideally provide a comprehensive examination of the targeted tissue, including the exact localization and metabolic activity of the target tissue, the tissue flow and functional changes within the surrounding tissues, and the pathognomonic changes leading to eventual disease. Due to the high effectiveness and wide application of multimodal imaging, it has been seen in the rapid evolution and becoming a standard practice in the clinic for diagnosing many diseases~\cite{moseley2004multimodality}.

The wide availability of multimodal medical images could greatly facilitate researchers in developing and validating advanced computer-aided medical diagnosis techniques. However, in practice tuples of corresponding images in different modalities data is very scarce and expensive to acquire due to the high cost and complex procedure for acquisition. The pressing need for data has even largely increased with the advent of deep neural networks, which are typically data-hungry and have become the standard approach in most machine learning tasks~\cite{goodfellow2016deep}. The limited amount and diversity of paired multimodal medical images could greatly restrict the capability of existing deep learning methods. Synthesizing multimodal images could be an attractive solution. Extensive studies have recently demonstrated that synthesizing medical images could successfully improve classification, registration and/or segmentation performance in medical imaging tasks~\cite{van2016,hu2019,lu2014,iglesias2013synthesizing,kim2018}. These studies proved that although synthetic data is derived from data that is already available, it can still improve performance of medical tasks as the synthesizer could learn a complete and continuous image distribution. Therefore, the ability to synthesize high-quality and clinical meaningful multimodal medical images is highly desirable, while such problem remains a widely unsolved challenge. Bi-modality image synthesis is a fundamental task towards the ultimate goal of synthesizing co-registered multimodal images. We focus our study on bi-modality image synthesis in this paper.

There exists a large amount of studies for medical image synthesis in the literature, ranging from conventional physical model-based methods~\cite{collins1998,hodneland2016} to the learning-based data-driven approaches~\cite{chen2017cross,costa2018,chartsias2018,li2018}. Recent studies mainly employ generative adversarial networks (GANs) for their fast advances and impressive results. However, most existing works focus on either single modal image synthesis~\cite{frangi2018} or cross-model image-to-image translation~\cite{nie2018}. The approaches for generating tuples of corresponding images in two modalities, from a common lowdimensional vector, are largely understudied. In the literature of natural image synthesis, there exists several methods~\cite{creswell2018,liu2016coupled,ghosh2018} for directly synthesizing multi-domain data from random noise vectors. Despite the success of these methods in handling natural image tasks, they can hardly achieve satisfactory results in the task of synthesizing bi-modality medical images due to the limited amount of bi-modality training data and high difficulties in capturing clinically meaningful characteristics of medical images.

In this work, we aim at a novel framework for generating high-quality, clinically meaningful bi-modality medical images using GANs. Specifically, the proposed framework consists of two generative modules to synthesize corresponding images in the two modalities in a sequential order. A complexity measurer is introduced to automatically determine the synthesis order. The modality that has a lower complexity is synthesized first via a decoder that maps a low dimensional latent vector to an image. And then the other modality with a higher complexity is synthesized later via an image-to-image translator which generates images with the guidance of previously generated counterparts in the other modality. The proposed synthesizer (i.e., the decoder and the image-to-image translator) is trained via semi-supervised learning. In supervised training, the synthesizer learns the joint distribution of bi-modality images explicitly by directly minimizing the reconstruction loss between fake and corresponding real images. To avoid overfitting to the limited paired bi-modality images, in unsupervised training the decoder and the image-to-image translator learns the marginal distribution for the two modalities respectively via adversarial learning. That is for each modality, we minimize the Wasserstein distance between distributions of the fake and real images in this modality. Despite learning marginal distributions in the unsupervised process, the joint distribution of bi-modality images is implicitly approximated by enforcing a weight sharing constraint on the first few deconvolutional layers of the decoder and the translator. As a result, the coarse spatial layouts of the corresponding images in the two modalities are decoded in the similar way based on the first few deconvolutional layers, and detailed and unique information in each modality is decoded differently based on the respective deconvolutional layers at the back. The resulting synthesizer could generate as many tuples of corresponding medical images in the two modalities as the user require, by sampling from a probability distribution that we impose to the associated latent space.

We take two task drivers, i.e., synthesizing apparent diffusion coefficient (ADC) maps and-T2 weight (T2w) image for prostate cancer classification, and T1w-T2w brain image synthesis, to comprehensively evaluate the proposed model and compare it with the state-of-the-art methods~\cite{costa2018,liu2016coupled,isola2017,zhu2017} based on four types of metrics: 1) Inception Score (IS)~\cite{salimans2016improved} and Fr\'echet Inception Distance (FID)~\cite{heusel2017gans}  which are widely-used metrics in the vision domain for evaluating the quality of synthetic images~\cite{lucic2018gans}; 2) the task-specific metric, i.e., classification accuracy, for assessing the effectiveness of synthetic images for clinically significant (CS) cancer vs. nonCS image classification; 3) Mutual Information Distance (MID) which measures the quality of multimodal synthetic images; and 4) a user study in which 3 radiologists with 1-year, 8year and 23-year reading experiences respectively are involved for assessing the clinically significance of synthetic multimodal images. Extensive experiments consistently demonstrate the superiority of our model to the state-of-the-art methods.

To summarize, the main contributions of this work include:

\begin{itemize}
    \item We present a novel bi-modality image synthesis network based on semi-supervised learning. Our network utilizes supervised training to ensure correct spatial correlation between synthetic images of the two modalities and meanwhile provides a high visual realism and diversity via unsupervised training. Extensive experimental results demonstrate the superiority of our method to the state-of-the-art methods~\cite{costa2018,liu2016coupled,isola2017,zhu2017}.
    \item We introduce a novel method that can effectively assess the complexity for synthesizing images of a modality. We integrate the complexity measurer into our image synthesis framework to automatically determine the synthesis order. The estimated complexity is consistent with both the experience and the final synthesis results.
    \item We conduct a very comprehensive evaluation based on two synthesis tasks, three types evaluation metrics and a user study to understand the performance of our method from various aspects, i.e., visual realism, diversity, correlations between corresponding bi-modality images, and clinical significance. In addition, we carefully analyze the impact of each proposed module, i.e., sequential architecture, semi-supervised learning and synthesis complexity measurer, and compare with the state-of-the-art methods via extensive experiments.
\end{itemize}
% % -----------------------------------------------------
\section{Related Work}
\label{sec:related}
% ------------------------------------------
Conventional method imaging synthesis mainly formulate a physical model of the observed data. These models can range from simple digital phantoms~\cite{collins1998} to more complex methodologies attempting to mimic anatomical and physiological medical knowledge~\cite{hodneland2016}. By modeling characteristics of the different acquisition devices, these methods can generate new high-quality images by sampling in an appropriate parameter space.

In recent years, we have witnessed an increasing popularity of the data-driven approaches for medical image synthesis. In this context, the underlying probability distribution that defines the manifold of real images is learned by machine learning techniques from a large pool of training images. Once trained, new images that lie on the manifold, i.e., realistic synthetic images, can be generated by sampling from the learned distribution. Various data-driven image synthesis approaches have been recently developed and successfully applied to medical applications. For instance, in~\cite{van2016,iglesias2013synthesizing,li2014deep,tang2019xlsor,tang2019deep} the authors demonstrated that synthesized data could improve the performance of classification and/or segmentation and/or registration of multimodal MRI images. In~\cite{nie2018,nie2017medical}, Nie et al. proposed a fully convolutional network with an adversarial loss and an image gradient difference loss to generate CT from an MRI image. However, these works main focus on single modal image synthesis or cross-model image-to-image translation. In this work, we aim at concurrently generating medical images of two modalities from common low dimensional vectors.

\begin{figure}[!t]
\centering
\includegraphics[width=0.48\textwidth]{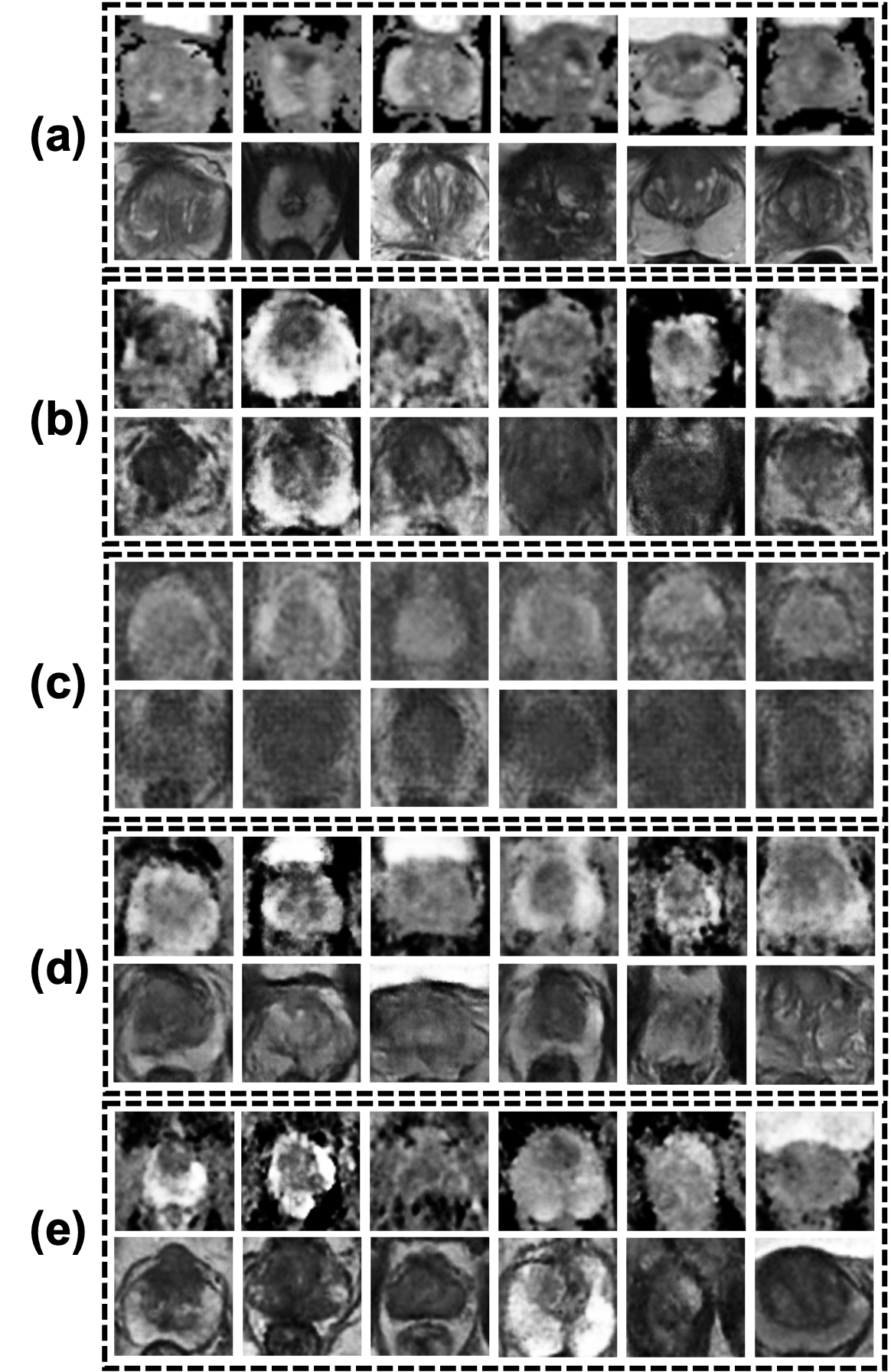}
\caption{Examples of six (a) real pairs of ADC-T2w images, and six synthetic ADC-T2w images based on (b) CoGAN~\cite{liu2016coupled}, (c) our sequential supervised GANs~\cite{costa2018}, (d) our sequential unsupervised GANs, (e) our sequential semi-supervised GANs. The top row of (a)–(e) shows the ADC maps and the bottom shows T2w.}
\label{fig1}
\end{figure}
\begin{figure}[thbp]
\centering
\includegraphics[width=0.48\textwidth]{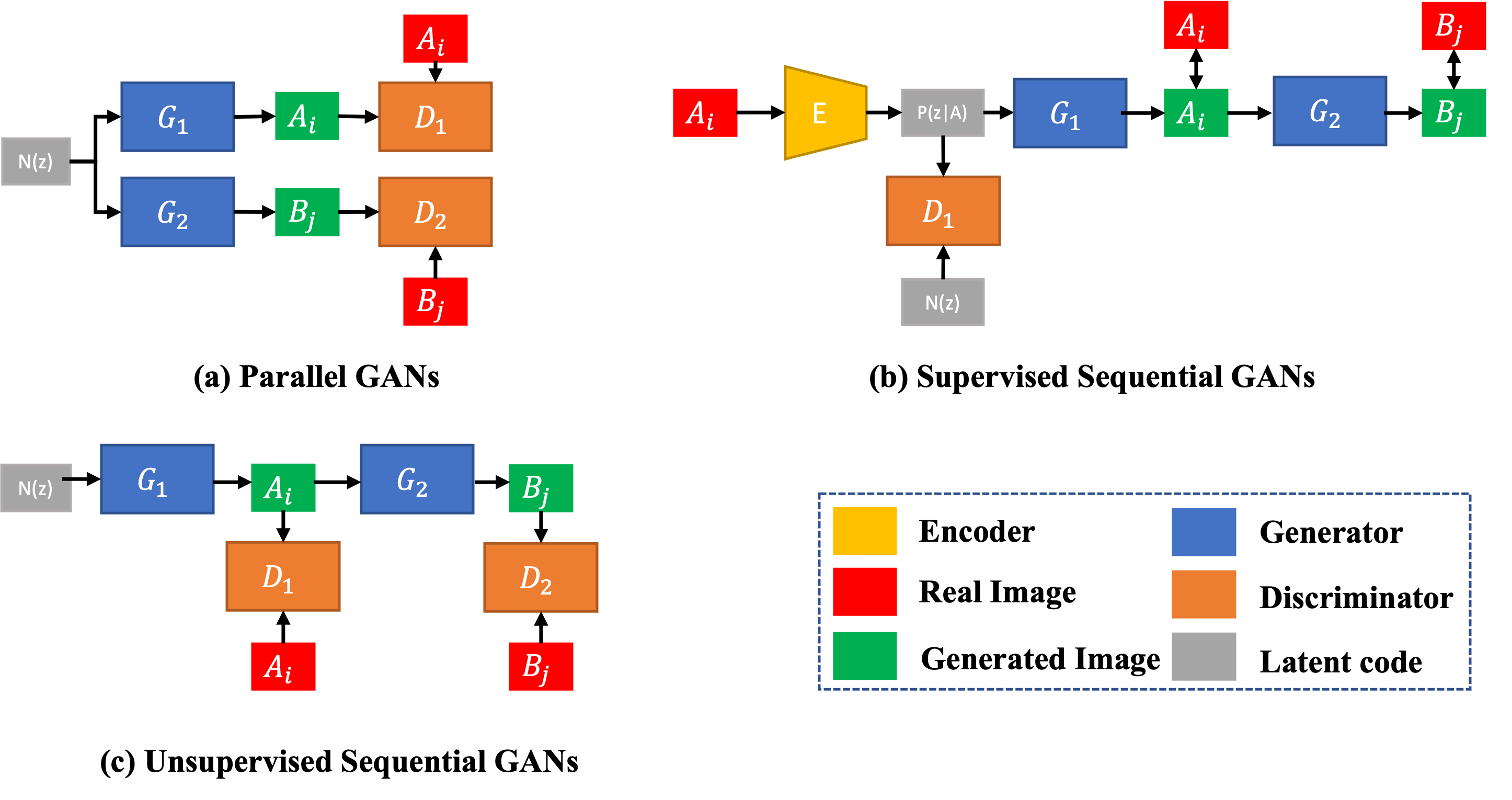}
\caption{Framework of (a) parallel GANs, (b) supervised sequential GANs, (c) unsupervised sequential GANs. In supervised training, paired bi-modality real images $A_i$ and $B_j$ are provided for training, where $A_i$ and $B_j$ denote images in two modalities and $i = j$. In unsupervised training, unpaired bi-modality real images $A_i$ and $B_j$ are provided for training and $i\ne j$.}
\label{fig2}
\end{figure}
In terms of multimodal image synthesis, existing methods can be categorized into two classes: 1) parallel multi-modal image synthesis which generates images of two modalities using two GANs from common low dimensional vectors, as shown in Fig.~\ref{fig2}(a), 2) sequential multi-modal image synthesis, which first generates images of one modality from low-dimensional vectors, followed by image-to-image translation that maps them to their counterparts in the other modality, as shown in Fig.~\ref{fig2}(b), (c).

Examples of the first class include~\cite{chartsias2018,liu2016coupled,jog2017random}. For instance, in~\cite{liu2016coupled} Liu and Tuzel proposed coupled GAN (CoGAN), in which two GANs were utilized to learn the marginal image distributions in different domains (e.g., color and depth images, face images with different attributes, etc.) respectively. Images of each domain are then synthesized by sampling from the corresponding marginal distribution. To guarantee the correct relationships between corresponding images in different domains, a weight-sharing constraint is enforced so that the several GANs decode high-level semantics in the same way. In \cite{chartsias2018} Chartsias et al. proposed a multimodal MR synthesis method by learning a shared modality-invariant latent space in which all input modalities are embedded. Their method was evaluated on two public datasets, i.e., ISLES~\cite{isles2015} and BRATS~\cite{brats2015}, to demonstrate its superiority to the state-of-the-art methods. However, these methods completely ignore the inherent differences of the task complexity arising from the acquisition characteristics of different modalities. For instance, as shown in Fig.~\ref{fig1}(a),the ADC map which captures the functional and coarse anatomical information of a prostate has a low spatial resolution. In comparison, the T2w image which captures the detailed anatomical structure of a prostate contains many high-frequency texture information. As a result, synthesizing T2w images of a prostate using parallel GANs, e.g., CoGAN, could be more challenging than synthesizing the corresponding ADC images due to the much higher complexity of synthesizing fine texture of T2w images. As a result, the quality of synthetic T2w images is much worse than that of ADC maps (as shown in Fig.~\ref{fig1}(b)).

To address the above issue, the second class of solutions, i.e., sequential bi-modality image synthesis, facilitates the consideration of different task complexities, providing a more flexible solution for multimodal medical image synthesis. In the sequential multimodal image synthesis framework, the modality with a lower synthesis complexity, is synthesized first. Images of the other modality are then generated via image-to-image translation. For instance, in \cite{costa2018} Costa et al. proposed to first generate a retinal vessel map via a GAN, followed by another GAN which synthesizes the corresponding retinal fundus image based on the synthetic vessel map. In \cite{costa2018}, the sequential GANs were trained in a supervised manner (as shown in Fig.~\ref{fig2}(b)), where the encodings (i.e., latent code) of real retinal images along with paired vessel maps are used. The constraint of the paired relationship between a vessel map and its corresponding retinal image is accomplished by minimizing both the reconstruction losses of vessel-fundus pairs and an adversarial loss. However, \cite{costa2018} requires tuples of corresponding images in different modalities/domains for training, which are challenging to obtain in practice. Limited training data makes the GANs only “see” very sparse distributed samples, yielding a great difficulty in modeling a complete and well-approximated joint distribution of true multimodal images and in turn lead to poor performance during the testing phase (as shown in Fig.~\ref{fig1}(c)). One potential solution to address this issue is to remove the correspondence dependency via unsupervised training, as shown in Fig.~\ref{fig2}(c). That is train the sequential GANs using unpaired multimodal images so that each GAN can learn the marginal distribution of a particular modality and in turn generate visually realistic images of this modality. However, unsupervised training cannot ensure correct relationships among different modalities (as shown in Fig.~\ref{fig1}(d)).
% % -----------------------------------------------------
\begin{figure*}[thbp]
\centering
\includegraphics[width=0.98\textwidth]{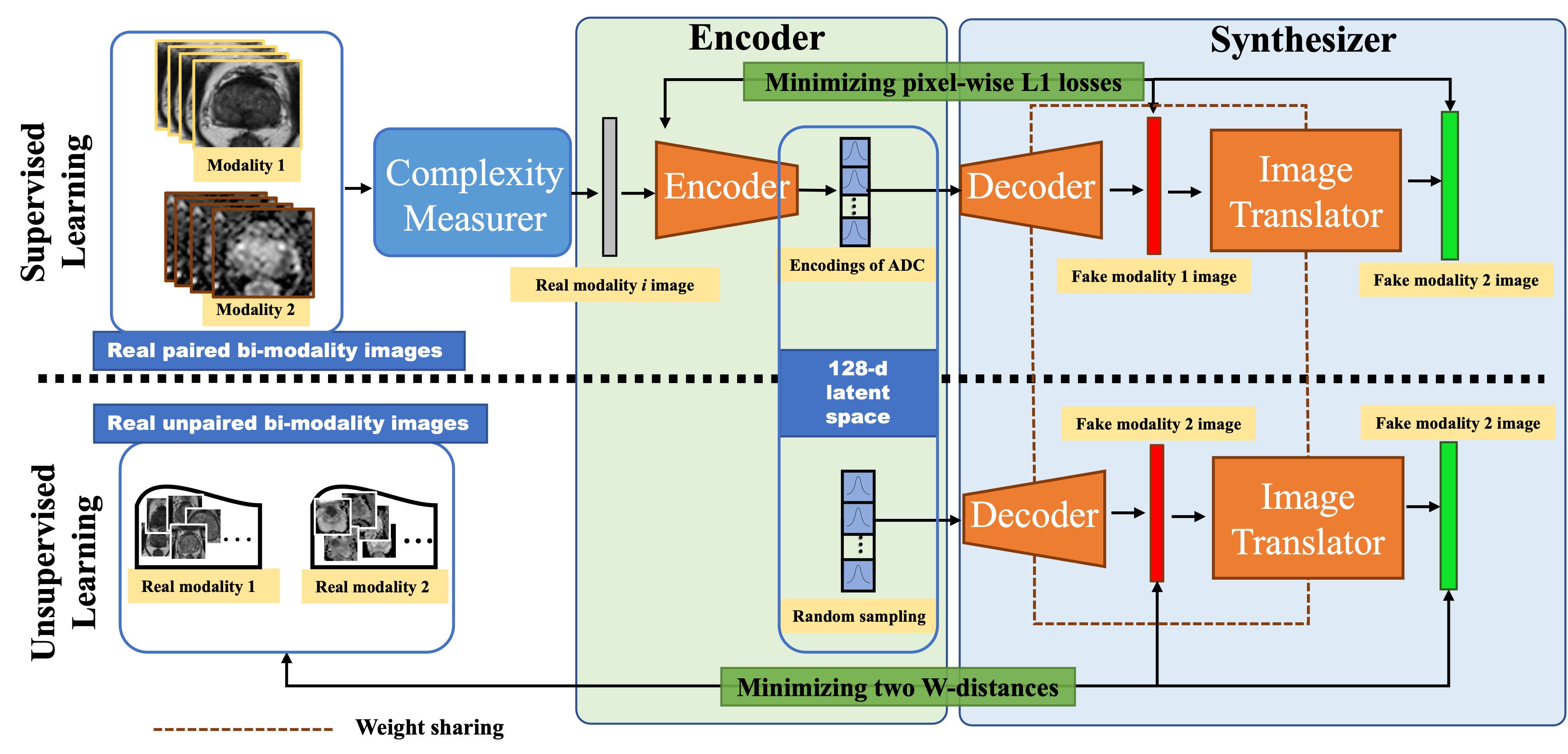}
\caption{Framework of our sequential semi-supervised GANs for bi-modality medical image synthesis.}
\label{fig3}
\end{figure*}
\section{Methodology}
In this study, our goal is to employ the generative adversarial learning technique to model the joint distribution of bi-modality medical images $(I^1_{syn},I^2_{syn})\sim\mathbb{P}_{syn}(I^1_{syn},I^2_{syn})$ based on a given set of training images $I^1_{real,m},I^2_{real,m}|m =1, 2,...,M$, where $I^k$ denotes the $k$-th modality, ($I^1_\bullet$, $I^2_\bullet$) indicates a tuple of spatially aligned bi-modality images, $M$ denotes the total number of training data. We desire that the manifold defined by the learned distribution $\mathbb{P}_{syn}(I^1_{syn},I^2_{syn})$ can well approximate the true manifold defined by $\mathbb{P}_{real}(I^1_{real},I^2_{real})$ so that synthetic images sampled from $\mathbb{P}_{syn}(I^1_{syn},I^2_{syn})$ are visually realistic and have a large diversity.

To achieve our goal, we propose a general bi-modality image synthesis framework based on sequential generative adversarial learning and semi-supervised learning, as shown in Fig.~\ref{fig3}. It mainly consists of three modules: 1) a complexity measurer which ranks the modalities in an increasing order of difficulty for synthesis ($i, j$); 2) an encoder which maps a real image of modality $i$ to a low dimensional latent vector $v_i$; and 3) a synthesizer which first decodes the latent vector $v_i$ to a fake image of modality $i$ and then translates image $I^i_{syn}$ to images of the modality $j$. The generator is trained using a semi-supervised manner. Our framework is for bi-modality medical image synthesis but it can be easily generalized to synthesize images of more than two modalities.

In the following, we first describe our bi-modality image synthesis model architecture, followed by details of semi-supervised training.

\subsection{Complexity Measurement for Sequential Synthesis}
In sequential synthesis, the modality easier to generate is synthesized earlier so that the difficulties of generating the other challenging modality can be greatly alleviated by conditioning on the prior information. How to determine the generation order so as to optimize the effectiveness of sequential synthesis is highly important; however, the method for quantitatively measure the synthesis complexity has not been studied to the best of our knowledge.

\begin{figure}[thbp]
\centering
\includegraphics[width=0.48\textwidth]{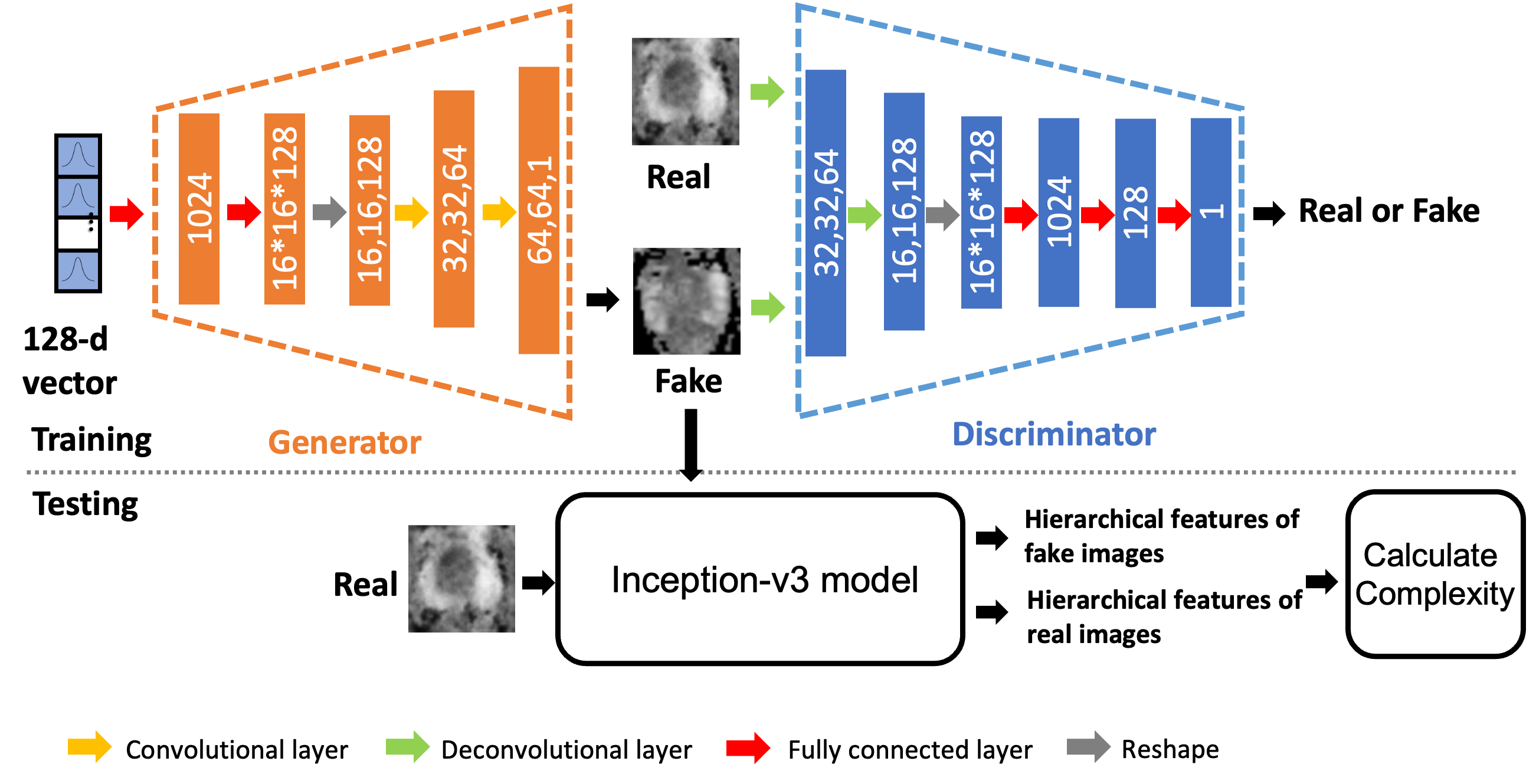}
\caption{Network architecture of the GAN-based complexity measurer.}
\label{fig4}
\end{figure}
In this paper, we propose to measure the complexity using a parallel GAN based on the quadratic Wasserstein-2 distance (W2-distance) in a hierarchical feature space. That is, we first synthesize $N$ images in each modality using a well-trained parallel GAN model, the architecture of each GAN is shown in Fig.~\ref{fig4}. The two GANs for two modalities respectively share identical network design (without weight sharing) and take a common 128-$d$ vector drawn from a predefined Gaussian distribution as input.

\begin{figure*}[thbp]
\centering
\includegraphics[width=0.98\textwidth]{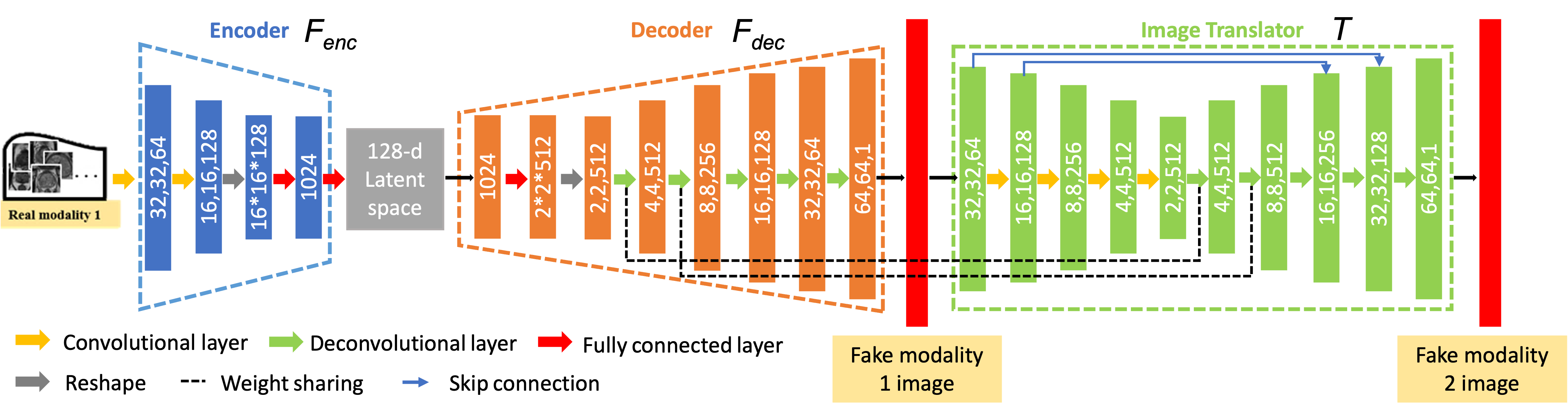}
\caption{Network architecture of the encoder $F_{\text{enc}}$ and the sequential generator (i.e., $F_{\text{dec}}$ and $T$).}
\label{fig5}
\end{figure*}
To measure the complexity of synthesizing two modalities, instead of directly calculating the W2-distance in the raw data space we follow the method in \cite{heusel2017gans} and do the calculation in a hierarchical feature space based on the pre-trained Inceptionv3 network~\cite{szegedy2016} on ImageNet~\cite{deng2009}  dataset to comprehensively measure the complexity in both shallow and in-depth visual features. Specifically, we model the feature distribution using a multivariate Gaussian distribution with mean and covariance. Accordingly, the distance between the distributions of the generated images' features and real images' features is calculated as Eq. (1). This distance reflects the generation complexity $C$ as for the modality easy to synthesize, it is more likely to correctly model the complete data distribution and yield a lower $C$ value in Eq. (1), and vice versa.
\begin{equation}
    C=  \sum _ {i=0}^{L} \left\{||\mu^i_{g}-\mu_{r}^{i}||^2_2   
+\text{Tr}\left( \Sigma_g^i +  \Sigma_r^i - 2\left(\Sigma_g^i\Sigma_r^i\right)^{\frac{1}{2}} \right)\right\}
\end{equation}
where $\mu^i$ and $\Sigma^i$ denote the mean and variance of a multivariate Gaussian distribution for features extracted using the $i$-th layer of the Inception-v3 model, $L$ denotes the total number of layers, the subscripts $g$ and $r$ indicate the generated and real data distributions respectively, $\text{Tr}$ sums up all the diagonal elements.

Note that ideally Inception-V3 model can be replaced by any CNN pretrained on a large dataset, here we choose Inception-V3 model because it has been widely demonstrated to be superior to the state-of-the-art methods~\cite{goodfellow2016deep,he2015} for its better accuracy, fewer parameters and faster speed, and the pretrained model on ImageNet is public available.

We calculate the generation complexity for ADC ($C_{\text{ADC}}$) and T2w ($C_{\text{T2w}}$) respectively based on 500 synthetic ADC images and 500 synthetic T2w images according to (1). $C_{\text{ADC}}$ and $C_{\text{T2w}}$ are 182.9 and 230.4 respectively, indicating a lower complexity of generating ADC images than T2w images. The result is consistent with the experience. Intuitively, compared with ADC, T2w provides more detailed texture information of the anatomical structure of a prostate, thus its more difficult for a synthesizer to correctly model the distribution of real T2w images than ADC images. Accordingly, the distance between the distributions of generated T2w images and real T2w images is usually greater than that between the distributions of generated ADC images and real ADC images.

\subsection{Semi-Supervised Training of the Sequential Synthesizer}
As shown in Fig.~\ref{fig3}, we train the sequential GAN in a semisupervised manner, with the goal that supervised learning can explicitly encode the relationships among different modalities in the image-to-image translator and unsupervised learning can correctly model the marginal distributions of the two modalities so that the synthetic multimodal images are visually realistic and have a large diversity. Fig.~\ref{fig5} shows the architecture of the encoder ($F_{\text{enc}}$) and the sequential synthesizer ($F_{\text{dec}}$ and $T$). $F_{\text{enc}}$ first encodes a real image of modality 1 into a 128 latent encoding via two convolutional layers, followed by a reshape layer and two fully connected layers. Then the 128 latent encoding is decoded into a fake image of modality 1 size of 64 × 64 via the $F_{\text{dec}}$. After that an autoencoder is applied as a translator $T$ to map the fake image of modality 1 into its counterpart image of modality 2. The first two deconvolutional layers of the $F_{\text{dec}}$ and $T$ share weights to ensure that the spatial layouts of the corresponding images of the two modalities are consistent.

In the following, we detail the training strategies used in our study.

\subsubsection{Supervised Training:}
The top part of Fig.~\ref{fig3} shows details of the supervised training process. We first utilize the encoder $F_\text{enc}$ to obtain encodings of real images of modality 1, i.e., $z = F_{\text{enc}}(I^1_a)$, where $I^1_a$ a denotes the $a$-th real image of modality 1. The decoder then reconstructs a fake image $\hat{I}^1_a = F_{\text{dec}}(z)$ based on $z$. After that, an image-to-image translator based on an autoencoder converts $\hat{I}^1_a$ a to a fake image of modality 2 as $\hat{I}^2_a=T(\hat{I}^1_a)$. In supervised training, paired multimodal images are provided and thus for each pair of fake images $\hat{I}^1_a,\hat{I}^2_a$ we could find their corresponding true pairs of images $\hat{I}^1_a,\hat{I}^2_a$. Therefore, we can train the entire network, including both $F_{\text{enc}}$ and the generator (i.e., $F_{\text{dec}}$ and $T$) by minimizing the pixel-wise reconstruction loss $L_1$ as:
\begin{equation}
      L_{1} = \mathbb{E}_{I^1,I^2 \sim p(I^{1},I^{2})} [||I^{1}-\hat{I^1}||+||I^{2}-\hat{I^2}||]
\end{equation}
where $\mathbb{E}_{I^1,I^2 \sim p(I^{1},I^{2})}$ is the expectation over pairs ($I^1 ,I^2$), sampled from the joint data distribution of real training pairs $p(I^1,I^2)$, $||x-\hat{x}||$ calculates the average of pixel-wise Manhattan distances between intensities of images $x$ and $\hat{x}$.

Ideally, the supervised training approach can make the generator (i.e., $F_\text{dec}$ and $T$) efficiently capture the correct paired relationships via explicitly minimizing the $L_1$ loss between synthetic and corresponding real data. However, only using supervised training approach would encounter a severe overfitting problem when the number of multimodal medical data is small. This is because the generator only 'sees' a very sparse and small portion of the latent space which contains encodings of real data in the training process. Consequently, during the testing process when synthesizing multimodal data from noise vectors $z \sim p(z)$ (where $p(z)$ conforms a Gaussian distribution), rather than encodings, the quality of the synthesized images could be extremely poor, even though we constrain the distribution of encodings to conform the same distribution. To address this problem, we also guide the generator to learn the marginal distribution of each image modality based on unpaired multimodal images via unsupervised learning.

\subsubsection{Unsupervised Training:}
The bottom part of Fig.~\ref{fig3} shows details of the unsupervised approach. Rather than training the generator using limited encodings from true images, the generator is trained using unlimited latent vectors drawn from $z \sim p(z)$ (where $p(z)$ conforms a Gaussian distribution) and unpaired multimodal images $P(I^1)$ and $P(I^2)$. We utilize two discriminators $D^1$ and $D^2$ to respectively approximate the W-distances $W^1$ and $W^2$ between the fake and real images as (3) for the two modalities respectively.
\begin{equation}
\begin{aligned}
    W^x = &\max \{\mathbb{E}_{I^x \sim \mathbb{P}_{\text{real}(I^x)}}[D^x(I^x)] \\
    &- \mathbb{E}_{I^x \sim \mathbb{P}_\text{syn}(\hat{I^x})}[D^x(I^x)] - \lambda_{I^x}R_{I^x} \} (x=1,2)
\end{aligned}
\end{equation}
where $I^x$ and $\hat{I^x} = F_\text{dec}(z)$ are real and synthetic images respectively, $R_{I^x}$ is used for enforcing the 1--Lipschitz constraint of $D^x(I^x)$, $\lambda_{I^x}$ is the parameter for adjusting the impact of $D^x(I^x)$~\cite{gulrajani2017improved}. Accordingly, the loss function of training the generator (i.e., $F_\text{dec}$ and $T$) is:
\begin{equation}
    L_{\text{unsup}} = W^1 + W^2
\end{equation}

Unsupervised training enables the generator focus on learning the marginal distribution of images of a single modality from unpaired multimodal images, leading to the generation of more visually realistic images of each modality. In addition, the weights of the first few layers of the decoder and the image-toimage translators are sharing, enabling the high-level semantic features between the corresponding images of the two modalities being decoded in the same way and in turn preserving inherent and coarse spatial correlations between images to different modalities. On the other hand, the weights of the last few layers are completely independent, capturing unique low-level detailed features of each modality.

\subsubsection{Semi-Supervised Training:}
To train the synthesizer in a semi-supervised manner, we first train it in a supervised manner by inputting a batch of paired training images (i.e., batch size is 32) and minimizing the $L_1$ loss defined in Eq. (2) using one iteration. Then we continue to train the decoder and image translator of the synthesizer in an unsupervised manner using another iteration by inputting a batch of unpaired-images (i.e., batch size is 32) and a 128-$d$ random vector of Gaussian distribution and minimizing the W-distance defined in Eq. (3). We alternate the supervised and unsupervised training processes for 40K iterations. Semi-supervised training enables both the correct paired relationships via supervised training, and a high visual realism and diversity via unsupervised training.
% % -----------------------------------------------------
\section{Experiments}
\subsection{Datasets and Applications}
In this study, we evaluate the proposed method using two medical tasks: ADC-T2w prostate image synthesis and T1wT2w brain image synthesis.

\subsubsection{Prostate Multi-Parametric Magnetic Resonance Imaging Dataset}
The study was approved by our local institutional review board. In this study, we utilized T2w and ADC of mp-MRI sequences for the task of image synthesis. An ADC map is derived from a transverse diffusion-weighted image (DWI) which provides pathological information of tissues. The derivation of ADC from DWI can be found in \cite{lim2009prostate}. The mp-MRI images are collected from two datasets: 1) a locally collected dataset including 156 patients data pathologically validated by a 12-core systematic TRUS-guided plus targeted prostate biopsy. Dataset details are listed in \cite{le2017automated,yang2017co}, 2) a public dataset PROSTATEx (training)~\cite{clark2013cancer,litjens2014,Litjens2015}, including data of 204 MRI-targeted biopsy-proven patients. Among 360 patients data, 226 patients are with benign prostatic hyperplasia (BPH) or indolent lesions, which are collectively referred to as non-Clinically Significant (CS) prostate cancer (PCa), and 134 patients are with CS PCa. From the two datasets, a radiologist manually selects 533 original ADC-T2w images containing CS PCa and 1992 ADC-T2w images containing nonCS PCa. The dataset was further divided into the training set (483 CS from 116 distinct cancerous patients) and the test set (50 CS from 10 distinct patients that are different from those in the training set) as \cite{wang2018stitchad}. We applied non-rigid registration~\cite{le2017automated} to every pair of ADC and T2w images to minimize motion-induced misalignment errors between the two modalities. Then, for each image pair, we manually cropped a rectangular region of interest size of 64 $\times$ 64 which encompasses the prostate area.

\subsubsection{IXI T1w-T2w Brain Dataset:}
The IXI dataset~\cite{ixi2005} consists of a variety of MR images from nearly 600 normal subjects from several hospitals in London. To make the numbers of training images and test images similar to those of the prostate dataset, we used 181 patients' data from the Hammersmith Hospital using a Philips 3T system and extracted two to three axial-plane T1w-T2w slices from each patient, yielding 533 co-registered T1w-T2w image pairs. The dataset was further divided into a training set (161 patients with 483 T1w-T2w image pairs) and a test set (20 patients with 50 T1w-T2w image pairs). We resized each of the 533 co-registered T1w-T2w image pairs from 256 $\times$ 256 to 128 $\times$ 128.

For both prostate and IXI dataset, we normalized the intensity of each image to [-1, 1].

\subsection{Evaluation Metrics and Implementation Details}
\subsubsection{Evaluation Metrics}
We evaluated the synthesis methods using three types of metrics: 1) Inception Score (IS)~\cite{salimans2016improved}  and Fr\'echet Inception Distance (FID)~\cite{heusel2017gans}  which are two widely used metrics in computer vision domain to assess the quality of single modal synthetic images; 2) the classification accuracy to assess the effectiveness of the synthetic images for training a classification model; 3) Mutual Information Distance (MID) which measures whether the correlations between corresponding synthesic images in different modalities are correct or not. In addition, we also conducted a user study to examine the quality and clinical significance of our synthetic images. In the following, we first explain the first three types metrics. Details of the user study will be presented in Sec.IV-D-2.

\noindent\textbf{Inception Score (IS):}
It is proposed in \cite{salimans2016improved} to simulate humans observations for determining the visual quality and diversity of synthetic images. Given a set of synthetic images $\{I_{\text{syn},1}, I_{\text{syn},2}, ...,  I_{\text{syn},N}\}$, we calculate conditional label distribution $p(y|I_{\text{syn},i})$ for each synthetic image $i$ based on the Inception model~\cite{szegedy2016}, where y denotes the classification label of the Inception model. For each $p(y|I_{\text{syn},i})$, we calculate its entropy to represent the image $I_{\text{syn},i}$'s visual quality. We also calculate the marginal distribution $p(y)$ by averaging conditional distributions of all synthetic samples and then compute the entropy of $p(y)$. A high entropy of $p(y)$ denotes a diverse distribution of $p(y|I_{\text{syn},i})$ among all images, reflecting a large diversity of the synthetic image set. By considering both the visual quality of each individual synthetic image and the diversity of the entire set, IS score is defined as:
\begin{equation}
\begin{aligned}
    \text{IS} &= \exp \left(\sum_1^N \text{KL}\left(p(y|I_{\text{syn}, i}) \| p(y)\right) \right) \\
    &=\exp \left(H(y) - \sum_1^N H(y|I_{\text{syn}, i}) \right)
\end{aligned}
\end{equation}
where $(x)$ represents entropy of variable $x$. A large IS denotes a high quality of the generated images.

\noindent\textbf{Fr\'echet Inception Distance (FID):}
It is proposed in \cite{heusel2017gans} to represent the realistic degree of synthetic images. To calculate FID, we map images into a common feature space by the Inception model and model the distributions of the synthetic and real features using two continuous multivariate Gaussian models respectively. Then the FID is computed by calculating the Fr\'echet distance between the two Gaussian models. A small FID denotes a high degree of reality of the generated images.

\noindent\textbf{Classification Accuracy:}
We desire that despite being generated from real data, synthetic images could effectively augment the size and diversity of real data. In this study, we evaluate the performance gain in the task of prostate cancer classification achieved by using synthetic images to train a clinically significant (CS) vs. non-clinically significant (nonCS) classifier. Specifically, we follow the network design of \cite{wang2018stitchad} for constructing our prostate CS vs. nonCS classifier. For multimodal data, we directly concatenate images of different modalities as input and apply the same network for classification. We train the classifier using 483 synthetic multimodal images and evaluate the classifier using 50 real CS images and 50 real nonCS images. All the testing data are neither used for training the classifier nor used for training the synthesizer.

\noindent\textbf{Mutual Information Distance (MID):}
In this study we define the metric Mutual Information Distance (MID),
\begin{equation}
    \text{MID} = \| \mathbb{E}_{(I^1_{\text{syn}},I^2_{\text{syn}})\sim P_{\text{syn}}} [\text{MI}_{\text{syn}}] 
    - \mathbb{E}_{(I^1_{\text{real}},I^2_{\text{real}})\sim P_{\text{real}}} [\text{MI}_{\text{real}}] \|
\end{equation}
which first computes the mutual information~\cite{maes1997} $\text{MI}_{\text{syn}}$ of synthetic image pairs $I^1_{\text{syn}}$ and $I^2_{\text{syn}}$ and the mutual information $\text{MI}_{\text{real}}$ of real image pairs $I^1_{\text{real}}$ and $I^2_{\text{real}}$, then calculates the absolute difference between $\text{MI}_{\text{syn}}$ and $\text{MI}_{\text{real}}$. Even though the absolute value of $\text{MI}_{\text{syn}}$ and $\text{MI}_{\text{real}}$ could be different for different images modalities, $\text{MI}_{\text{syn}}$ should be close to $\text{MI}_{\text{real}}$ if synthetic image pairs are well aligned with each other, and vice verses. Thus, we believe MID can well reflect the correctness of spatial relationship between synthetic images of different modalities. A small MID denotes a high quality of the synthetic multimodal images in terms of correctly encoding the true correlations between images of different modalities.

\subsubsection{Implementation Details:}
All the models were trained using the Adam optimizer with an initial learning rate of 1e-4 and the batch size of 32. We trained each model for 40K iterations. For training the classification network, the initial learning rate is 0.01, the factor of learning rate decays to 0.99 for every 30 steps, the batch size is 64, the optimizer is SGD with a weight decay of 1e-4. We trained the classifier for 10K iterations. Our method was implemented in Python with TensorFlow, and all the experiments were conducted on a single NVidia Titan X GPU.

To obtain a statistical result of the IS, FID and MID scores, we equally divided 500 synthetic data into 10 groups. We calculated the IS, FID and MID scores for each synthetic group, and then computed the mean and standard deviation of the scores. To measure the classification accuracy, we trained five classifiers using 483 synthetic positive samples and 483 real negative samples, and tested the classifiers on 50 real positive samples and 50 real negative samples.

\begin{table}[]
    \centering
    \caption{COMPARISON OF DIFFERENT STRUCTURES OF GANSFOR SYNTHESIZING PROSTATE ADC-T2W IMAGES}
    \setlength{\tabcolsep}{3pt}{
    \begin{tabular}{ccccc}
    \toprule
    Modality & Evaluation & Parallel GANs & \multicolumn{2}{c}{Sequential GANs} \\
    \cmidrule{4-5}
    & Metrics & & z-ADC-T2w & z-T2w-ADC \\
    \midrule
    \multirow{2}{*}{ADC} & IS & 1.77$\pm$0.07 & 1.72$\pm$0.07 & \textbf{1.98$\pm$0.19} \\
    & FID & 312.05$\pm$7.02 & 306.05$\pm$10.06 & \textbf{268.73$\pm$7.90} \\
    \midrule
    \multirow{2}{*}{T2w} & IS & 1.80$\pm$0.06 & \textbf{2.22$\pm$0.12} & 2.01$\pm$0.12 \\
    & FID & 309.80$\pm$6.13 & \textbf{254.42$\pm$7.88} & 280.17$\pm$8.53 \\
    \midrule
    ADC & Class. Acc. (\%) & 86.20$\pm$0.40 & \textbf{86.50$\pm$0.92} & 82.90$\pm$0.30 \\
    T2w & Class. Acc. (\%) & 83.40$\pm$0.80 & \textbf{83.50$\pm$0.50} & 80.40$\pm$0.66 \\
    \bottomrule
    \end{tabular}}
    \label{tab1}
\end{table}

\begin{table}[]
    \centering
    \caption{COMPARISON OF DIFFERENT STRUCTURES OF GANS FOR SYNTHESIZING BRAIN T1W-T2W IMAGES}
    \setlength{\tabcolsep}{4pt}{
    \begin{tabular}{ccccc}
    \toprule
    Modality & Evaluation & Parallel GANs & \multicolumn{2}{c}{Sequential GANs} \\
    \cmidrule{4-5}
    & Metrics & & z-T1w-T2w & z-T2w-ADC \\
    \midrule
    \multirow{2}{*}{T1w} & IS & \textbf{1.50$\pm$0.10} & 1 .35$\pm$0.07 & 1.39$\pm$0.10 \\
    & FID & 134.84$\pm$3.43 & \textbf{131.62$\pm$6.06} & 138.18$\pm$8.86 \\
    \midrule
    \multirow{2}{*}{T2w} & IS & 1.21$\pm$0.03 & \textbf{1.23$\pm$0.05} & 1.22$\pm$0.05 \\
    & FID & 93.75$\pm$2.27 & \textbf{69.11$\pm$4.42} &  96.49$\pm$2.92 \\
    \bottomrule
    \end{tabular}}
    \label{tab2}
\end{table}

\begin{table}[]
    \centering
    \caption{COMPARISON OF DIFFERENT TRAINING STRATEGIES FOR SYNTHESIZING PROSTATE ADC-T2W IMAGES}
    \setlength{\tabcolsep}{4pt}{
    \begin{tabular}{ccccc}
    \toprule
    Modality & Eva. Metrics & Supervised & Unsupervised & Semi-supervised \\
    \midrule
    \multirow{2}{*}{T1w} & IS & 1.59$\pm$0.09 & 1.91$\pm$0.07 & \textbf{1.95$\pm$0.11} \\
    & FID & 300.06$\pm$4.50 & 289.40$\pm$5.92 & \textbf{263.78$\pm$8.21} \\
    \midrule
    \multirow{2}{*}{T2w} & IS & 1.62$\pm$0.07 & 2.21$\pm$0.33 & \textbf{2.61$\pm$0.24} \\
    & FID & 333.05$\pm$10.32 & 245.15$\pm$25.87 & \textbf{234.89$\pm$9.62} \\
    \midrule
    \multirow{3}{*}{ADC-T2w} & FID & 239.05$\pm$7.94 & 192.76$\pm$19.42 & \textbf{179.54$\pm$5.38} \\
    & Acc. (\%) & 59.50$\pm$2.84 & 90.20$\pm$0.40 & \textbf{93.00$\pm$0.45} \\
    & MID & 0.024$\pm$0.004 & 0.025$\pm$0.004 & \textbf{0.011$\pm$0.006} \\
    \bottomrule
    \end{tabular}}
    \label{tab3}
\end{table}

\subsection{Ablation Study}
In this subsection, we exam three factors which affect the performance of a GAN-based bi-modality image synthesis method: 1) parallel GANs vs. sequential GANs, 2) the order of the synthesis tasks, and 3) training strategies, i.e., supervised, unsupervised or semi-supervised. For all the following experiments, we synthesize 500 pairs of bi-modality images.

\subsubsection{Parallel GANs vs. Sequential GANs:}
In this experiment we compare the performance of two structures, i.e.,  parallel and sequential, for synthesizing bi-modality images in two medical tasks, i.e.,  prostate ADC-T2w image pairs and T1w-T2w brain images. To eliminate the effects arising from the training strategies, for both structures we adopt the unsupervised training strategy and the two network structures are illustrated in Fig.~\ref{fig2}(a) and (c). For the sequential GANs, we experiment with two different orders for both tasks, i.e., z-ADC-T2w and z-T2w-ADC for prostate data and z-T1w-T2w and z-T2w-T1w for brain data, where z denotes a 128-$d$imensional noise vector randomly selected from a fixed Gaussian distribution. Table~\ref{tab1} reports the results of synthetic images based on different network structures. The first four rows compare the quality of each single modality using the metrics of IS and FID. Clearly, for all metrics and modalities sequential GANs achieves superior performance to the parallel GANs. The last two rows compare the accuracy of the classifiers which are trained using synthetic images generated from different GANs and tested using real images. For all cases, the classifiers trained using synthetic images from sequential GANs outperforms those based on parallel GANs.

\begin{figure}[thbp]
\centering
\includegraphics[width=0.48\textwidth]{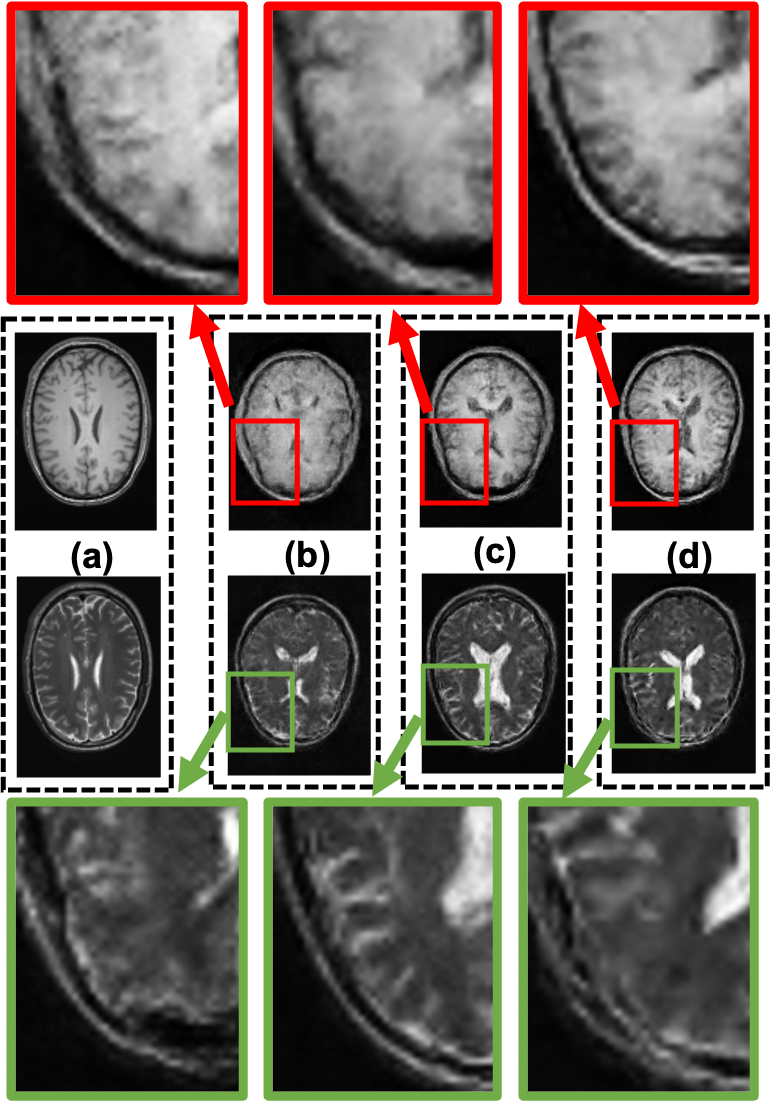}
\caption{Examples of (a) a real pair of T1w-T2w images, synthetic T1w-T2w images based on (b) parallel unsupervised GANs, (c) sequential unsupervised GANs (z-T1w-T2w), (d) sequential unsupervised GANs (z-T2w-T1w). The top row of (a)–(d) shows the T1w images and the bottom row shows the T2w images. The visualization results show that: 1) sequential GAN outperforms parallel GAN for generating images of better quality for one modality and similar quality for the other modality; 2) images of the modality synthesized later in the sequential GAN has a better visual quality than the other, i.e., T2w in (c) and T1w in (d) are more visually close to real images than the other.}
\label{fig6}
\end{figure}
Table~\ref{tab2} compares the performance of the two structures for synthesizing T1w-T2w brain images. Similarly, the performance of sequential GANs is consistently better for most metrics and modalities except for T1w using the IS metric. As the IXI dataset does not provide specific classification task (i.e., all data is from normal patients), we omit the classification accuracy in Table~\ref{tab2}. Fig.~\ref{fig6} shows a set of real T1w-T2w brain images and synthetic image pairs based on a parallel GAN and sequential GANs of two different orders. In general, the sequential GAN outperforms the parallel GAN for generating images of better quality for one modality and similar quality for the other.

\subsubsection{Complexity Measurement for Sequential Synthesis:}
We further validate the effectiveness of the complexity measurement for sequential synthesis. According to the method described in Sec. III-A) , the complexity of synthesizing ADC images and T2w images are 182.9 and 230.4 respectively, indicating a greater difficulty in generating T2w images than ADC images. The last two columns of Table~\ref{tab1} compare the quality of synthetic bi-modality images using different orders. Results show that generally speaking for most cases, the generation order z-ADCT2w achieves superior performance to the order z-T2w-ADC. These results validate two points: 1) the proposed method is effective in reflecting the complexity of a synthesis task, and 2) performing easier synthesis task first could achieve better performance for the bi-modality image synthesis task.

Similarly, we also exam the complexity of synthesizing T1w and T2w brain images which are very close to each other, indicating a similar difficulty in generating T1w and T2w images. By comparing the last two columns of Table~\ref{tab2} we also observe that the IS and FID of the images generated based on two different orders are similar, which are consistent with our previous conclusions.

\subsubsection{Comparison of Different Training Strategies:}
Table~\ref{tab3} compares different training strategies on the task of synthesizing ADC-T2w prostate images. Our training set contains 483 paired ADC-T2w images with biopsy proven prostate cancers, the size of the images is 64 $\times$ 64. For the supervised approach, we used all the 483 paired ADC-T2w images for training. For the unsupervised method, we randomly shuffled the paring relation of ADC and T2w images to form 483 unpaired ADC-T2w pairs for training. As for the semi-supervised method, we alternately trained the network in a supervised manner using the 483 paired ADC-T2w images and unsupervised manner using the 483 unpaired images. We believe the comparison is fair as identical 483 ADC and T2w images are utilized for training for all the three approaches. For each training strategy, we synthesize 500 ADC-T2w image pairs from 128-$d$ noise vectors for the evaluation.

The first four rows report the quality of the synthetic images of each modality. The results in the fifth row are obtained by concatenating each pair of synthetic ADC-T2w images and input it into the network for calculating the FID score. This score reflects the distance between the joint distributions of features extracted from synthetic and real ADC-T2w images. The sixth row denotes the accuracy of the classifier trained based on the synthetic images and the last row shows the MID score which reflects the mutual information inconsistency between synthetic and real multimodal images. The results show that the unsupervised approach outperforms the supervised approach and semi-supervised training achieves the best performance for all metrics than the other two strategies. This is understandable as the supervised model is prone to overfit to limited encodings from the real training images and in turn perform poorly for synthesizing images from unseen 128-$d$ noise vectors. In comparison, the unsupervised model takes 128-$d$ noise vectors as input for training, and thus it sees a much large variety of input latent codes and in turn well learns the marginal distribution of each single modality. Semi-supervised training can capture both correct spatial correlation via supervised training and learn the marginal distribution of each single modality and thus can provide better visual quality, greater diversity and more accurate spatial consistency than the other two methods. In addition, we also consider the classifier trained based on 483 real positive ADC-T2w data and 483 real negative ADC-T2w data as a baseline. The baseline classifier achieves the accuracy of 93.40 $\pm$ 0.40\%, which is very close to that achieved by our synthetic images, demonstrating a similar effectiveness of our synthetic images as real images for training a classifier.

\begin{table*}[]
    \centering
    \caption{COMPARISON OF THE STATE OF THE-ARTS ON PROSTATE ADC-T2W IMAGE SYNTHESIS}
    \setlength{\tabcolsep}{12pt}{
    \begin{tabular}{ccccccc}
    \toprule
    Modality & Eva. Metrics & Costa et al.~\cite{costa2018} & CoGAN~\cite{liu2016coupled} & CycleGAN~\cite{zhu2017} & pix2pix~\cite{isola2017} & Ours \\
    \midrule
    \multirow{2}{*}{ADC} & IS & 1.53$\pm$0.05 & 1.79$\pm$0.16 & -- & -- & \textbf{1.95$\pm$0.11} \\
    & FID & 360.83$\pm$6.66 & 299.71$\pm$12.74 & -- & -- & \textbf{263.78$\pm$8.21} \\
    \midrule
    \multirow{2}{*}{T2w} & IS & 1.60$\pm$0.07 & 1.68$\pm$0.09 & 1,.76$\pm$0.09 & 1.67$\pm$0.11 & \textbf{2.61$\pm$0.24} \\
    & FID & 342.00$\pm$7.88 & 302.14$\pm$6.79 & 293.46$\pm$9.69 & 373.99$\pm$5.85 & \textbf{234.89$\pm$9.62} \\
    \midrule
    \multirow{3}{*}{ADC-T2w} & FID & 237.44$\pm$1.95 & 240.32$\pm$9.54 & -- & -- & \textbf{179.54$\pm$5.38} \\
    & Acc. (\%) & 53.90$\pm$2.39 & 88.90$\pm$0.70 & -- & -- & \textbf{93.00$\pm$ 0.45} \\
    & MID & 0.025$\pm$0.002 & 0.017$\pm$0.006 & -- & -- & \textbf{0.011$\pm$0.006} \\
    \bottomrule
    \end{tabular}}
    \label{tab4}
\end{table*}
\begin{table}[]
    \centering
    \caption{RESULTS OF THE USER STUDY PERFORMED BY THREE RADIOLOGISTS ($R_1$, $R_2$, AND $R_3$)}
    \setlength{\tabcolsep}{6pt}{
    \begin{tabular}{cccccc}
    \toprule
    Tests & Method & Evaluation Items & $R_1$ & $R_2$ & $R_3$ \\
    \midrule
    \multirow{3}{*}{Test I} & Real & TPR (\%) & 67.8 & 93.4 & 41.6 \\
    & CoGAN & FPR (\%) & 0.6 & 28.2 & 5.8 \\
    & Ours & FPR (\%) & \textbf{4.4} & \textbf{91.4} & \textbf{7.8} \\
    \midrule
    \multirow{3}{*}{Test II} & Real & Avg. Score & 2.39 & 2.34 & 2.47 \\
     & CoGAN & Avg. Score & 1.05 & 1.96 & 1.08 \\
     & Ours & Avg. Score & \textbf{1.19} & \textbf{2.21} & \textbf{1.39} \\
    \bottomrule
    \end{tabular}}
    \label{tab5}
\end{table}

\subsection{Comparison With the State-of-the-Art}
\subsubsection{Quantitative Evaluation}
We compare our method with four state-of-the-art GAN-based image synthesis methods using the prostate image synthesis task, including Costa et al.'s method~\cite{costa2018}, CoGAN~\cite{liu2016coupled}, CycleGAN~\cite{zhu2017} and pix2pix~\cite{isola2017}. As described in Sec. II, Costa et al.'s method also employs a sequential architecture which first utilizes a GAN to synthesize a retinal vessel tree and then converts the generated tree into a retinal image via the U-Net~\cite{ronneberger2015u}. The synthesizer of \cite{costa2018} is trained in a supervised manner. In this experiment, we adopted the same order z-ADC-T2w as our method to first synthesize ADC images and then map them to the corresponding T2w images for \cite{costa2018}. CoGAN~\cite{liu2016coupled} is the state-of-the-art model trained in an unsupervised manner for synthesizing multi-domain images. CoGAN adopts a parallel structure that consists of two parallel GANs to concurrently synthesize images of two modalities. In CoGAN, spatial correlation between two modalities are enforced by weight sharing between the decoders of the two GANs. CycleGAN and pix2pix are the state-of-the-art models for translating images of one modality to another modality. Both CycleGAN and pix2pix only learn to map encodings from real images of one modality to images of the other. As a result, given limited training images which is a common scenario in medical domain, the two methods cannot learn a dense mapping from one modality to another modality and in turn usually lead to blurry synthesis results in the test phase. Compared with those four methods, our method employs semi-supervised training which enables our method to well approximate the joint distribution of the two modalities. In addition, we propose a task complexity measurement approach which can automatically determine the synthesis order to ensure an optimized performance.

Results in Table~\ref{tab4} show that our method outperforms all other comparison methods for all cases. As CycleGAN and pix2pix can only synthesize one modality based on the real input of the other modality, we take real ADC images as input and generate fake T2w images as the corresponding counterparts. Table~\ref{tab4} reports only the results of FID and IS for synthetic T2w images for these two methods. Fig.~\ref{fig1} displays some exemplar synthetic multimodal images by our method, CoGAN (which achieves the second-best performance among the comparison methods) as well as the real data. As shown in Fig.~\ref{fig1}, the visual quality of the synthetic T2w images by CoGAN is obviously worse than ours due to the ignorance of task complexity differences.

\subsubsection{User Study}
To further judge the visual realism and clinical significance of our results from the perspective of radiologists, we use perceptual experiments with human observes, as proposed in \cite{salimans2016improved,zhu2017toward,chen2017photographic}. Different from these approaches, we focus on medical image synthesis. We asked 3 radiologists with 1 year, 8 years and 23 years clinical experience, to evaluate our synthesized prostate ADC-T2w images. Each radiologist reads approximate 120 MRI scans per day.

The user study consists of 2 tests.

\noindent\textbf{Test I:}
In each round, one pair of ADC-T2w images (randomly sampled from 500 real data, 500 synthetic data based on our method and 500 synthetic data based on CoGAN) is presented to a radiologist. The radiologist is asked to identify whether the presented ADC-T2w images is real or not. If a synthetic pair is labeled as real, it is considered as a false positive (FP). A high false positive rate (FPR) denotes greater clinical significance and visual realism to the radiologist as the corresponding synthetic images can fool the radiologists. In addition, we also recorded the true positive rate (TPR), i.e.,  the ratio of real pairs that are also labeled as real to the total number of real samples. The TPR value can represent a radiologist's tendency for labeling uncertain samples. A high TPR and a high FPR denote the radiologist tends to label uncertain samples as real while a low value for both TPR and FPR denotes that he/she tends to label uncertain samples as fake.

\noindent\textbf{Test II:}
In each round, we present 3 pairs of ADC-T2w images, including one real ADC-T2w data, one synthetic data based on our method and one synthetic data based on CoGAN. Each pair is randomly selected from 500 samples, and the presentation order of the three pairs is random in each round. The radiologist is asked to score three pairs from 1 to 3 according to their visual realism and clinical significance (1 means the worst quality and 3 means the best quality).

Compared with Test I, Test II imposes higher challenges on synthetic images as it could be easier for a radiologist to distinguish real from fake by exhibiting both real and fake data in a common window for comparison. Each test (i.e.,  Test I and Test II) was performed by all the three radiologists. Table~\ref{tab5} shows the average results of the two tests. For Test I, the FPR results achieved by CoGAN are consistently lower than those achieved by Ours for all radiologists, denoting that the radiologists are more likely to consider our results as real images. Therefore, the synthetic images from our method are more visually realistic and of greater clinical significance than those from CoGANs.

In Test II, we average scores for synthetic images of each category (i.e., real, Ours, CoGAN). The results show that the scores achieved by our method are 5.6\% $\sim$ 50.2\% lower than those of real data, indicating that there remains differences between synthetic images and real images to radiologists. These differences mainly due to the clarity of synthetic T2w images compared with real data and the ambiguity of the shape of prostate glands, peripheral zones, central zones and bladder in synthetic images. When comparing the results of Ours and CoGANs, the average score of Ours is 5.0\% $\sim$ 22.3\% higher than that of CoGAN's, demonstrating the superiority of our method to CoGAN for bimodality medical image synthesis.

% % -----------------------------------------------------
 \section{Conclusion}
In this work, a sequential generative model is presented for synthesizing spatially-aligned pairs of 2D images in two modalities. The complexity of synthesizing images in each modality is assessed and the one with a lower complexity is synthesized first, followed by the one with a higher complexity generated via an image-to-image translator. The sequential generative model learns the joint distribution of multimodal images via semi-supervised learning. Extensive experimental results demonstrate that our model can generate visually realistic and clinically meaningful bi-modality medical images. The synthetic images can effectively augment training data and in turn greatly improve the accuracy of the CNN-based classifier. In addition to classification, our synthetic images which contain simulated lesions should be also beneficial to lesion detection and segmentation tasks. There are several studies in the literature exploring weakly-supervised CNN for prostate cancer detection and segmentation from mp-MRI images~\cite{yang2017co,wang2018,yang2017joint}. That is, training a CNN-based detector or segmenter using images containing image-level labels, i.e., cancerous vs. noncancerous. By augmenting positive training images using our synthetic prostate ADC-T2w images, we could make the detector or segmenter 'see' more possible cancer-related visual features and in turn improve the performance of the weakly-supervised CNN. We will investigate this in our future work. Our future work also includes extending the 2D multimodal image synthesizer to 3D, and synthesizing images of more than two modalities.

% -----------------------------------------------------
\bibliographystyle{IEEEtran.bst}
\bibliography{refs}

% Generated by IEEEtran.bst, version: 1.12 (2007/01/11)
\begin{thebibliography}{10}
\providecommand{\url}[1]{#1}
\csname url@samestyle\endcsname
\providecommand{\newblock}{\relax}
\providecommand{\bibinfo}[2]{#2}
\providecommand{\BIBentrySTDinterwordspacing}{\spaceskip=0pt\relax}
\providecommand{\BIBentryALTinterwordstretchfactor}{4}
\providecommand{\BIBentryALTinterwordspacing}{\spaceskip=\fontdimen2\font plus
\BIBentryALTinterwordstretchfactor\fontdimen3\font minus
  \fontdimen4\font\relax}
\providecommand{\BIBforeignlanguage}[2]{{%
\expandafter\ifx\csname l@#1\endcsname\relax
\typeout{** WARNING: IEEEtran.bst: No hyphenation pattern has been}%
\typeout{** loaded for the language `#1'. Using the pattern for}%
\typeout{** the default language instead.}%
\else
\language=\csname l@#1\endcsname
\fi
#2}}
\providecommand{\BIBdecl}{\relax}
\BIBdecl

\bibitem{moseley2004multimodality}
M.~Moseley and G.~Donnan, ``Multimodality imaging: introduction,''
  \emph{Stroke}, vol.~35, no. 11\_suppl\_1, pp. 2632--2634, 2004.

\bibitem{goodfellow2016deep}
I.~Goodfellow, Y.~Bengio, A.~Courville \emph{et~al.}, ``Deep learning: From
  basics to practice, vol. 1,'' 2016.

\bibitem{van2016}
G.~van Tulder and M.~de~Bruijne, ``Combining generative and discriminative
  representation learning for lung ct analysis with convolutional restricted
  boltzmann machines,'' \emph{IEEE Transactions on Medical Imaging}, vol.~35,
  no.~5, pp. 1262--1272, 2016.

\bibitem{hu2019}
B.~Hu, Y.~Tang, E.~I.-C. Chang, Y.~Fan, M.~Lai, and Y.~Xu, ``Unsupervised
  learning for cell-level visual representation in histopathology images with
  generative adversarial networks,'' \emph{IEEE Journal of Biomedical and
  Health Informatics}, vol.~23, no.~3, pp. 1316--1328, 2019.

\bibitem{lu2014}
C.~Lu and M.~Mandal, ``Toward automatic mitotic cell detection and segmentation
  in multispectral histopathological images,'' \emph{IEEE Journal of Biomedical
  and Health Informatics}, vol.~18, no.~2, pp. 594--605, 2014.

\bibitem{iglesias2013synthesizing}
J.~E. Iglesias, E.~Konukoglu, D.~Zikic, B.~Glocker, K.~Van~Leemput, and
  B.~Fischl, ``Is synthesizing mri contrast useful for inter-modality
  analysis?'' in \emph{Medical Image Computing and Computer-Assisted
  Intervention--MICCAI 2013: 16th International Conference, Nagoya, Japan,
  September 22-26, 2013, Proceedings, Part I 16}.\hskip 1em plus 0.5em minus
  0.4em\relax Springer, 2013, pp. 631--638.

\bibitem{kim2018}
K.~Kim and H.~Myung, ``Autoencoder-combined generative adversarial networks for
  synthetic image data generation and detection of jellyfish swarm,''
  \emph{IEEE Access}, vol.~6, pp. 54\,207--54\,214, 2018.

\bibitem{collins1998}
D.~Collins, A.~Zijdenbos, V.~Kollokian, J.~Sled, N.~Kabani, C.~Holmes, and
  A.~Evans, ``Design and construction of a realistic digital brain phantom,''
  \emph{IEEE Transactions on Medical Imaging}, vol.~17, no.~3, pp. 463--468,
  1998.

\bibitem{hodneland2016}
E.~Hodneland, E.~Hanson, A.~Z. Munthe-Kaas, A.~Lundervold, and J.~M.
  Nordbotten, ``Physical models for simulation and reconstruction of human
  tissue deformation fields in dynamic mri,'' \emph{IEEE Transactions on
  Biomedical Engineering}, vol.~63, no.~10, pp. 2200--2210, 2016.

\bibitem{chen2017cross}
M.~Chen, A.~Carass, A.~Jog, J.~Lee, S.~Roy, and J.~L. Prince, ``Cross contrast
  multi-channel image registration using image synthesis for mr brain images,''
  \emph{Medical image analysis}, vol.~36, pp. 2--14, 2017.

\bibitem{costa2018}
P.~Costa, A.~Galdran, M.~I. Meyer, M.~Niemeijer, M.~Abràmoff, A.~M. Mendonça,
  and A.~Campilho, ``End-to-end adversarial retinal image synthesis,''
  \emph{IEEE Transactions on Medical Imaging}, vol.~37, no.~3, pp. 781--791,
  2018.

\bibitem{chartsias2018}
A.~Chartsias, T.~Joyce, M.~V. Giuffrida, and S.~A. Tsaftaris, ``Multimodal mr
  synthesis via modality-invariant latent representation,'' \emph{IEEE
  Transactions on Medical Imaging}, vol.~37, no.~3, pp. 803--814, 2018.

\bibitem{li2018}
N.~Li, Z.~Zheng, S.~Zhang, Z.~Yu, H.~Zheng, and B.~Zheng, ``The synthesis of
  unpaired underwater images using a multistyle generative adversarial
  network,'' \emph{IEEE Access}, vol.~6, pp. 54\,241--54\,257, 2018.

\bibitem{frangi2018}
A.~F. Frangi, S.~A. Tsaftaris, and J.~L. Prince, ``Simulation and synthesis in
  medical imaging,'' \emph{IEEE Transactions on Medical Imaging}, vol.~37,
  no.~3, pp. 673--679, 2018.

\bibitem{nie2018}
D.~Nie, R.~Trullo, J.~Lian, L.~Wang, C.~Petitjean, S.~Ruan, Q.~Wang, and
  D.~Shen, ``Medical image synthesis with deep convolutional adversarial
  networks,'' \emph{IEEE Transactions on Biomedical Engineering}, vol.~65,
  no.~12, pp. 2720--2730, 2018.

\bibitem{creswell2018}
A.~Creswell, T.~White, V.~Dumoulin, K.~Arulkumaran, B.~Sengupta, and A.~A.
  Bharath, ``Generative adversarial networks: An overview,'' \emph{IEEE Signal
  Processing Magazine}, vol.~35, no.~1, pp. 53--65, 2018.

\bibitem{liu2016coupled}
M.-Y. Liu and O.~Tuzel, ``Coupled generative adversarial networks,''
  \emph{Advances in neural information processing systems}, vol.~29, 2016.

\bibitem{ghosh2018}
A.~Ghosh, V.~Kulharia, V.~Namboodiri, P.~H. Torr, and P.~K. Dokania,
  ``Multi-agent diverse generative adversarial networks,'' in \emph{2018
  IEEE/CVF Conference on Computer Vision and Pattern Recognition}, 2018, pp.
  8513--8521.

\bibitem{isola2017}
P.~Isola, J.-Y. Zhu, T.~Zhou, and A.~A. Efros, ``Image-to-image translation
  with conditional adversarial networks,'' in \emph{2017 IEEE Conference on
  Computer Vision and Pattern Recognition (CVPR)}, 2017, pp. 5967--5976.

\bibitem{zhu2017}
J.-Y. Zhu, T.~Park, P.~Isola, and A.~A. Efros, ``Unpaired image-to-image
  translation using cycle-consistent adversarial networks,'' in \emph{2017 IEEE
  International Conference on Computer Vision (ICCV)}, 2017, pp. 2242--2251.

\bibitem{salimans2016improved}
T.~Salimans, I.~Goodfellow, W.~Zaremba, V.~Cheung, A.~Radford, and X.~Chen,
  ``Improved techniques for training gans,'' \emph{Advances in neural
  information processing systems}, vol.~29, 2016.

\bibitem{heusel2017gans}
M.~Heusel, H.~Ramsauer, T.~Unterthiner, B.~Nessler, and S.~Hochreiter, ``Gans
  trained by a two time-scale update rule converge to a local nash
  equilibrium,'' \emph{Advances in neural information processing systems},
  vol.~30, 2017.

\bibitem{lucic2018gans}
M.~Lucic, K.~Kurach, M.~Michalski, S.~Gelly, and O.~Bousquet, ``Are gans
  created equal? a large-scale study,'' \emph{Advances in neural information
  processing systems}, vol.~31, 2018.

\bibitem{li2014deep}
R.~Li, W.~Zhang, H.-I. Suk, L.~Wang, J.~Li, D.~Shen, and S.~Ji, ``Deep learning
  based imaging data completion for improved brain disease diagnosis,'' in
  \emph{Medical Image Computing and Computer-Assisted Intervention--MICCAI
  2014: 17th International Conference, Boston, MA, USA, September 14-18, 2014,
  Proceedings, Part III 17}.\hskip 1em plus 0.5em minus 0.4em\relax Springer,
  2014, pp. 305--312.

\bibitem{tang2019xlsor}
Y.-B. Tang, Y.-X. Tang, J.~Xiao, and R.~M. Summers, ``Xlsor: A robust and
  accurate lung segmentor on chest x-rays using criss-cross attention and
  customized radiorealistic abnormalities generation,'' in \emph{International
  Conference on Medical Imaging with Deep Learning}.\hskip 1em plus 0.5em minus
  0.4em\relax PMLR, 2019, pp. 457--467.

\bibitem{tang2019deep}
Y.-X. Tang, Y.-B. Tang, M.~Han, J.~Xiao, and R.~M. Summers, ``Deep adversarial
  one-class learning for normal and abnormal chest radiograph classification,''
  in \emph{Medical Imaging 2019: Computer-Aided Diagnosis}, vol. 10950.\hskip
  1em plus 0.5em minus 0.4em\relax SPIE, 2019, pp. 305--311.

\bibitem{nie2017medical}
D.~Nie, R.~Trullo, J.~Lian, C.~Petitjean, S.~Ruan, Q.~Wang, and D.~Shen,
  ``Medical image synthesis with context-aware generative adversarial
  networks,'' in \emph{Medical Image Computing and Computer Assisted
  Intervention- MICCAI 2017: 20th International Conference, Quebec City, QC,
  Canada, September 11-13, 2017, Proceedings, Part III 20}.\hskip 1em plus
  0.5em minus 0.4em\relax Springer, 2017, pp. 417--425.

\bibitem{jog2017random}
A.~Jog, A.~Carass, S.~Roy, D.~L. Pham, and J.~L. Prince, ``Random forest
  regression for magnetic resonance image synthesis,'' \emph{Medical image
  analysis}, vol.~35, pp. 475--488, 2017.

\bibitem{isles2015}
``Ischemic stroke lesion segmentation (isles) 2015 challenge,''
  \url{http://www.isles-challenge.org/ISLES2015}, 2015.

\bibitem{brats2015}
``Brain tumour segmentation (brats) 2015 challenge,''
  \url{https://sites.google.com/site/braintumorsegmentation/home/ brats2015},
  2015.

\bibitem{szegedy2016}
C.~Szegedy, V.~Vanhoucke, S.~Ioffe, J.~Shlens, and Z.~Wojna, ``Rethinking the
  inception architecture for computer vision,'' in \emph{2016 IEEE Conference
  on Computer Vision and Pattern Recognition (CVPR)}, 2016, pp. 2818--2826.

\bibitem{deng2009}
J.~Deng, W.~Dong, R.~Socher, L.-J. Li, K.~Li, and L.~Fei-Fei, ``Imagenet: A
  large-scale hierarchical image database,'' in \emph{2009 IEEE Conference on
  Computer Vision and Pattern Recognition}, 2009, pp. 248--255.

\bibitem{he2015}
K.~He, X.~Zhang, S.~Ren, and J.~Sun, ``Delving deep into rectifiers: Surpassing
  human-level performance on imagenet classification,'' in \emph{2015 IEEE
  International Conference on Computer Vision (ICCV)}, 2015, pp. 1026--1034.

\bibitem{gulrajani2017improved}
I.~Gulrajani, F.~Ahmed, M.~Arjovsky, V.~Dumoulin, and A.~C. Courville,
  ``Improved training of wasserstein gans,'' \emph{Advances in neural
  information processing systems}, vol.~30, 2017.

\bibitem{lim2009prostate}
H.~K. Lim, J.~K. Kim, K.~A. Kim, and K.-S. Cho, ``Prostate cancer: apparent
  diffusion coefficient map with t2-weighted images for detection—a
  multireader study,'' \emph{Radiology}, vol. 250, no.~1, pp. 145--151, 2009.

\bibitem{le2017automated}
M.~H. Le, J.~Chen, L.~Wang, Z.~Wang, W.~Liu, K.-T.~T. Cheng, and X.~Yang,
  ``Automated diagnosis of prostate cancer in multi-parametric mri based on
  multimodal convolutional neural networks,'' \emph{Physics in Medicine \&
  Biology}, vol.~62, no.~16, p. 6497, 2017.

\bibitem{yang2017co}
X.~Yang, C.~Liu, Z.~Wang, J.~Yang, H.~Le~Min, L.~Wang, and K.-T.~T. Cheng,
  ``Co-trained convolutional neural networks for automated detection of
  prostate cancer in multi-parametric mri,'' \emph{Medical image analysis},
  vol.~42, pp. 212--227, 2017.

\bibitem{clark2013cancer}
K.~Clark, B.~Vendt, K.~Smith, J.~Freymann, J.~Kirby, P.~Koppel, S.~Moore,
  S.~Phillips, D.~Maffitt, M.~Pringle \emph{et~al.}, ``The cancer imaging
  archive (tcia): maintaining and operating a public information repository,''
  \emph{Journal of digital imaging}, vol.~26, pp. 1045--1057, 2013.

\bibitem{litjens2014}
G.~Litjens, O.~Debats, J.~Barentsz, N.~Karssemeijer, and H.~Huisman,
  ``Computer-aided detection of prostate cancer in mri,'' \emph{IEEE
  Transactions on Medical Imaging}, vol.~33, no.~5, pp. 1083--1092, 2014.

\bibitem{Litjens2015}
------, ``Prostatex challenge data,''
  \url{https://doi.org/10.7937/K9TCIA.2017.MURS5CL}, 2017.

\bibitem{wang2018stitchad}
Z.~Wang, Y.~Lin, C.~Liao, K.-T. Cheng, and X.~Yang, ``Stitchad-gan for
  synthesizing apparent diffusion coefficient images of clinically significant
  prostate cancer.'' in \emph{BMVC}, 2018, p. 240.

\bibitem{ixi2005}
``Ixi dataset,'' \url{https://brain-development.org/ixi-dataset}, 2005.

\bibitem{maes1997}
F.~Maes, A.~Collignon, D.~Vandermeulen, G.~Marchal, and P.~Suetens,
  ``Multimodality image registration by maximization of mutual information,''
  \emph{IEEE Transactions on Medical Imaging}, vol.~16, no.~2, pp. 187--198,
  1997.

\bibitem{ronneberger2015u}
O.~Ronneberger, P.~Fischer, and T.~Brox, ``U-net: Convolutional networks for
  biomedical image segmentation,'' in \emph{Medical Image Computing and
  Computer-Assisted Intervention--MICCAI 2015: 18th International Conference,
  Munich, Germany, October 5-9, 2015, Proceedings, Part III 18}.\hskip 1em plus
  0.5em minus 0.4em\relax Springer, 2015, pp. 234--241.

\bibitem{zhu2017toward}
J.-Y. Zhu, R.~Zhang, D.~Pathak, T.~Darrell, A.~A. Efros, O.~Wang, and
  E.~Shechtman, ``Toward multimodal image-to-image translation,''
  \emph{Advances in neural information processing systems}, vol.~30, 2017.

\bibitem{chen2017photographic}
Q.~Chen and V.~Koltun, ``Photographic image synthesis with cascaded refinement
  networks,'' in \emph{2017 IEEE International Conference on Computer Vision
  (ICCV)}, 2017, pp. 1520--1529.

\bibitem{wang2018}
Z.~Wang, C.~Liu, D.~Cheng, L.~Wang, X.~Yang, and K.-T. Cheng, ``Automated
  detection of clinically significant prostate cancer in mp-mri images based on
  an end-to-end deep neural network,'' \emph{IEEE Transactions on Medical
  Imaging}, vol.~37, no.~5, pp. 1127--1139, 2018.

\bibitem{yang2017joint}
X.~Yang, Z.~Wang, C.~Liu, H.~M. Le, J.~Chen, K.-T. Cheng, and L.~Wang, ``Joint
  detection and diagnosis of prostate cancer in multi-parametric mri based on
  multimodal convolutional neural networks,'' in \emph{Medical Image Computing
  and Computer Assisted Intervention- MICCAI 2017: 20th International
  Conference, Quebec City, QC, Canada, September 11-13, 2017, Proceedings, Part
  III 20}.\hskip 1em plus 0.5em minus 0.4em\relax Springer, 2017, pp. 426--434.

\end{thebibliography}

\end{document}